# Fast and Adaptive Sparse Precision Matrix Estimation in High Dimensions

Weidong Liu[a], Xi Luo[b,c,d,*]

[a] *Department of Mathematics and Institute of Natural Sciences, Shanghai Jiao Tong University, Shanghai, CHINA*
[b] *Department of Biostatistics and Center for Statistical Sciences, Brown University, Providence, Rhode Island, USA*
[c] *Brown Institute for Brain Science, Brown University, Providence, Rhode Island, USA*
[d] *Initiative for Computation in Brain and Mind, Brown University, Providence, Rhode Island, USA*

**Abstract**

This paper proposes a new method for estimating sparse precision matrices in the high dimensional setting. It has been popular to study fast computation and adaptive procedures for this problem. We propose a novel approach, called Sparse Column-wise Inverse Operator, to address these two issues. We analyze an adaptive procedure based on cross validation, and establish its convergence rate under the Frobenius norm. The convergence rates under other matrix norms are also established. This method also enjoys the advantage of fast computation for large-scale problems, via a coordinate descent algorithm. Numerical merits are illustrated using both simulated and real datasets. In particular, it performs favorably on an HIV brain tissue dataset and an ADHD resting-state fMRI dataset.

## 1. Introduction

Estimating covariance matrices is fundamental in multivariate analysis. It has been popular to estimate the inverse covariance (or precision) matrix in the high dimensional setting, where the number of variables $p$ goes to infinity with the sample size $n$ (more precisely, in this paper, $p \gg n$ and $(\log p)/n = o(1)$). Inverting the sample covariance matrix has been known to be unstable for estimating the precision matrix. Recent proposals usually formulate this objective as regularized/penalized optimization problems, where regularization is employed to control the sparsity of the precision matrix. Besides the challenge of solving such large optimization problems, there is an important issue on how to choose an appropriate regularization level that is adaptive to the data. To address these two challenges, we propose a fast and adaptive method, and establish the theoretical properties when the regularization level is chosen by cross validation.

*Corresponding author
*Email address:* xi.rossi.luo@gmail.com, Phone: +1 401 863 6321 (Xi Luo)

Let $\boldsymbol{X} = (X_1, \ldots, X_p)^T$ be a $p$-variate random vector with a covariance matrix $\boldsymbol{\Sigma}$ or its corresponding precision matrix $\boldsymbol{\Omega} := \boldsymbol{\Sigma}^{-1}$. Suppose we observe independent and identically distributed random samples $\{\boldsymbol{X}_1, \ldots, \boldsymbol{X}_n\}$ from the distribution of $\boldsymbol{X}$. To encourage a sparse and stable estimate for $\boldsymbol{\Omega}$, regularized/penalized likelihood approaches have been proposed. Here, sparsity means that most of the entries in $\boldsymbol{\Omega}$ are exactly zero. Popular penalties include the $\ell_1$ penalty [1] and its extensions, for example, [2], [3], [4], and [5]. In particular, [3] developed an efficient algorithm, glasso, to compute the penalized likelihood estimator, and its convergence rates were obtained under the Frobenius norm [5] and the elementwise $\ell_\infty$ norm and spectral norm [6]. Other penalties were also studied before. For example, the $\ell_1$ penalty was replaced by the nonconvex SCAD penalty [7, 8, 9]. Due to the complexity of the penalized likelihood objective, theoretical analysis and computation are rather involved. Moreover, the theory usually relies on some theoretical assumptions of the penalty, and thus it provides limited guidance for applications.

Recently, column-wise or neighborhood based procedures have caught much attention, due to the advantages in both computation and analysis. [10] proposed to recover the support of $\boldsymbol{\Omega}$ using $\ell_1$ penalized regression, aka LASSO [1], in a row by row fashion. This can be computed efficiently via path-following coordinate descent [11] for example. A Dantzig selector proposal, replacing the LASSO approach, was proposed recently by [12], and the computation is based on standard solvers for linear programming. [13] proposed a procedure, CLIME, which seeks a sparse precision matrix under a matrix inversion constraint. Their procedure is also solved column by column via linear programming. Compared with the regularized likelihood approaches, their convergence rates were obtained under several matrix norms mentioned before, without imposing the mutual incoherence condition [6], and were improved when $\boldsymbol{X}$ follows polynomial tail distributions. However, all these procedures are computational expensive for very large $p$, and again these estimators were analyzed based on theoretical choices of the penalty.

Cross validation on the other hand has gained popularity for choosing the penalty levels or tuning parameters, because it is adaptive and usually yields superior performance in practice. Unfortunately, the theoretical understanding of cross validation is sparse. For a related problem on estimating sparse covariance matrices, [14] analyzed the performance of covariance thresholding where the threshold is based on cross validation. [15] provided a different approach using self-adaptive thresholding. However, these covariance estimation results cannot be extended to the inverse covariance setting, partly due to the problem complexity. This paper will provide theoretical justification for cross validation when estimating the precision matrix. This result is made possible because we propose a new column-wise procedure that is easy to compute and analyze. To the best of our knowledge, this paper is among the first to provide theoretical justification of cross validation for sparse precision matrix estimation.

The contributions of this paper are several folds. First, we propose a novel and penalized column-wise procedure, called Sparse Columnwise Inverse Operator (SCIO), for estimating the precision matrix $\boldsymbol{\Omega}$. Second, we establish the theoretical justification under mild conditions when its penalty is chosen by cross validation. The theory for cross validation is summarized as follows. A matrix is called $s_p$-sparse if there are at most $s_p$ non-zero elements on each row. It is shown that the error between our cross validated estimator $\hat{\boldsymbol{\Omega}}$ and $\boldsymbol{\Omega}$ satisfies $\|\hat{\boldsymbol{\Omega}}^1 - \boldsymbol{\Omega}\|_F^2/p = O_P(s_p(\log p)/n)$, where $\|\cdot\|_F$ is the Frobenius norm. Third, theoretical guarantees for the SCIO estimator are also obtained under other matrix norms, for example the element-wise $\ell_\infty$ norm which achieves graphical model selection [16]. Fourth, we provide a fast and simple algorithm for computing the estimator. Because our algorithm exploits the advantages of conjugate gradient and coordinate descent, and thus it provides superior performance in computational speed and cost. In particular, we reduce two nested loops in glasso [3] to only one. An R package of our method, *scio*, has been developed, and is publicly available on CRAN.

The rest of the paper is organized as follows. In Section 2, after basic notations and definitions are introduced, we present the SCIO estimator. Finite sample convergence rates are established with the penalty level chosen both by theory in Section 3 and by cross validation in Section 4. The algorithm for solving SCIO is introduced in Section 5. Its numerical merits are illustrated using simulated and real datasets. Further discussions on the connections and differences of our results with other related work are given in Section 6. The supplementary material includes additional results for the numerical examples in Section 5 and the proof of the main results.

The notations in this paper are collected here. Throughout, for a vector $\mathbf{a} = (a_1, \ldots, a_p)^T \in I\!R^p$, define $|\mathbf{a}|_1 = \sum_{j=1}^p |a_j|$ and $|\mathbf{a}|_2 = \sqrt{\sum_{j=1}^p a_j^2}$. All vectors are column vectors. For a matrix $\boldsymbol{A} = (a_{ij}) \in I\!R^{p\times q}$, we define the elementwise $l_\infty$ norm $|\boldsymbol{A}|_\infty = \max_{1\le i\le p, 1\le j\le q} |a_{ij}|$, the spectral norm $\|\boldsymbol{A}\|_2 = \sup_{|\mathbf{x}|_2\le 1} |\boldsymbol{A}\mathbf{x}|_2$, the matrix $\ell_1$ norm $\|\boldsymbol{A}\|_{L_1} = \max_{1\le j\le q} \sum_{i=1}^p |a_{ij}|$, the matrix $\infty$ norm $\|\boldsymbol{A}\|_\infty = \max_{1\le i\le q} \sum_{j=1}^p |a_{ij}|$, the Frobenius norm $\|\boldsymbol{A}\|_F = \sqrt{\sum_{i,j} a_{ij}^2}$, and the elementwise $\ell_1$ norm $\|\boldsymbol{A}\|_1 = \sum_{i=1}^p \sum_{j=1}^q |a_{ij}|$. $\boldsymbol{A}_{i,\cdot}$ and $\boldsymbol{A}_{\cdot,j}$ denote the $i$th row and $j$th column respectively. $\boldsymbol{I}$ denotes an identity matrix. $1\{\cdot\}$ is the indicator function. The transpose of $\boldsymbol{A}$ is denoted by $\boldsymbol{A}^T$. For any two matrices $\boldsymbol{A}$ and $\boldsymbol{B}$ of proper sizes, $\langle \boldsymbol{A}, \boldsymbol{B}\rangle = \sum_i \left(\boldsymbol{A}^T\boldsymbol{B}\right)_{ii}$. For any two index sets $T$ and $T^{'}$ and a matrix $\boldsymbol{A}$, we use $\boldsymbol{A}_{TT'}$ to denote the $|T|\times|T^{'}|$ matrix with rows and columns of $\boldsymbol{A}$ indexed by $T$ and $T^{'}$ respectively. The notation $\boldsymbol{A}\succ 0$ means that $\boldsymbol{A}$ is positive definite. For two real sequences $\{a_n\}$ and $\{b_n\}$, write $a_n = O(b_n)$ if there exists a constant $C$ such that $|a_n| \le C|b_n|$ holds for large

$n$, $a_n = o(b_n)$ if $\lim_{n\to\infty} a_n/b_n = 0$, and $a_n \asymp b_n$ if $a_n = O(b_n)$ and $b_n = O(a_n)$. Write $a_n = O_P(b_n)$ if $a_n = O(b_n)$ holds with the probability going to 1. The constants $C, C_0, C_1, \ldots$ may represent different values at each appearance.

## 2. Methodology

Our estimator is motivated by adding the $\ell_1$ penalty [1] to a column loss function, which is related to conjugate descent and a constrained minimization approach CLIME [13]. The technical derivations that lead to the estimator is provided in the supplementary material. Denote the sample covariance matrix by $\hat{\boldsymbol{\Sigma}}$. Let a vector $\hat{\boldsymbol{\beta}}_i$ be the solution to the following equation:

$$\hat{\boldsymbol{\beta}}_i = \arg\min_{\boldsymbol{\beta}\in \mathbb{R}^p} \left\{ \frac{1}{2}\boldsymbol{\beta}^T \hat{\boldsymbol{\Sigma}}\boldsymbol{\beta} - \boldsymbol{e}_i^T \boldsymbol{\beta} + \lambda_{ni}|\boldsymbol{\beta}|_1 \right\}, \tag{1}$$

where $\hat{\boldsymbol{\beta}}_i = (\hat{\beta}_{i1}, \ldots, \hat{\beta}_{ip})^T$, $\boldsymbol{e}_i$ is the $i$th column of a $p \times p$ identity matrix, and $\lambda_{ni} > 0$ is a tuning parameter. The tuning parameter could be different from column to column, adapting to different magnitude and sparsity of each column.

One can formulate a precision matrix estimate where each column is the corresponding $\hat{\boldsymbol{\beta}}_i$. However, the resulting matrix may not be symmetric. Similar to a symmetrization step employed in CLIME, we define the SCIO estimator $\hat{\boldsymbol{\Omega}} = (\hat{\omega}_{ij})_{p\times p}$, using the following symmetrization step,

$$\hat{\omega}_{ij} = \hat{\omega}_{ji} = \hat{\beta}_{ij} 1\{|\hat{\beta}_{ij}| < |\hat{\beta}_{ji}|\} + \hat{\beta}_{ji} 1\{|\hat{\beta}_{ij}| \geq |\hat{\beta}_{ji}|\}. \tag{2}$$

As we will establish in Section 3, similar to the results of CLIME, the convergence rates shall not change if the diagonal of the sample covariance $\hat{\boldsymbol{\Sigma}}$ is added by a small positive amount, as long as in the order of $n^{-1/2} \log^{1/2} p$. With this modification, (1) is then strictly convex and has a unique solution. In Section 5, we will present an efficient coordinate descent algorithm to solve it.

The SCIO estimator, like other penalized estimators, depends on the choice of $\lambda_{ni}$. We allow $\lambda_{ni}$ to be different from column to column, so that it is possible to adapt to each column's magnitude and sparsity, as we will illustrate in Section 4. More importantly, due to the simplified column loss function (1), we are able to establish, in Section 4, the theoretical guarantees when $\lambda_{ni}$ is chosen by cross validation. In comparison, the theory of cross validation for glasso [3] and CLIME [13] has not been established before, to the best of our knowledge.

## 3. Theoretical guarantees

### *3.1. Conditions*

Let $\mathcal{S}_i$ be the support of $\boldsymbol{\Omega}_{\cdot,i}$, the $i$th column of $\boldsymbol{\Omega} = (\omega_{ij})_{p\times p}$. Define the $s_p$-sparse matrices class

$$\mathcal{U} = \Big\{\boldsymbol{\Omega} \succ 0 : \max_{1\le j\le p} \sum_{i=1}^{p} 1\{\omega_{ij} \neq 0\} \le s_p, \quad \|\boldsymbol{\Omega}\|_{L_1} \le M_p, \\ c_0^{-1} \le \Lambda_{\min}(\boldsymbol{\Omega}) \le \Lambda_{\max}(\boldsymbol{\Omega}) \le c_0\Big\},$$

where $c_0$ is a positive constant, $\Lambda_{\min}(\boldsymbol{\Omega})$ and $\Lambda_{\max}(\boldsymbol{\Omega})$ are the minimum and maximum eigenvalues of $\boldsymbol{\Omega}$ respectively. The sparsity $s_p$ is allowed to grow with $p$, as long as it satisfies the following condition.

**(C1).** Suppose that $\boldsymbol{\Omega} \in \mathcal{U}$ with

$$s_p = o\left(\sqrt{\frac{n}{\log p}}\right) \tag{3}$$

and

$$\max_{1\le i\le p} \left\| \boldsymbol{\Sigma}_{\mathcal{S}_i^c \mathcal{S}_i} \left(\boldsymbol{\Sigma}_{\mathcal{S}_i \mathcal{S}_i}\right)^{-1} \right\|_\infty \le 1-\alpha \tag{4}$$

for some $0 < \alpha < 1$.

As we will see from Theorem 1, condition (3) is required for proving the consistency. Condition (4) is in the same spirit as the mutual incoherence or irrepresentable condition for glasso [6], but it is slightly relaxed, see Remark 2. In general, this type of conditions is believed to be almost necessary for penalization methods to recover support.

Let $\boldsymbol{Y} = (Y_1, \ldots, Y_p)^T = \boldsymbol{\Omega X} - \boldsymbol{\Omega \mu}$ where $\boldsymbol{\mu} = \mathsf{E}\boldsymbol{X}$. The covariance matrix of $\boldsymbol{Y}$ is thus $\boldsymbol{\Omega}$. The second condition is on the moments of $\boldsymbol{X}$ and $\boldsymbol{Y}$.

**(C2).** (Exponential-type tails) Suppose that $\log p = o(n)$. There exist positive numbers $\eta > 0$ and $K > 0$ such that

$$\mathsf{E}\exp\left(\eta(X_i-\mu_i)^2\right) \le K, \quad \mathsf{E}\exp\left(\eta Y_i^2\right) \le K \quad \text{for all } 1 \le i \le p.$$

**(C2*).** (Polynomial-type tails) Suppose that for some $\gamma, c_1 > 0$, $p \le c_1 n^{\gamma}$, and for some $\delta > 0$

$$\mathsf{E}|X_i-\mu_i|^{4\gamma+4+\delta} \le K, \quad \mathsf{E}|Y_i|^{4\gamma+4+\delta} \le K \quad \text{for all } i.$$

We will assume either one of these two types of tails in our main analysis. These two conditions are standard for analyzing sparse precision matrix estimation, see [13] and references within.

*3.2. Convergence rates of* $\hat{\mathbf{\Omega}} - \mathbf{\Omega}$

The first theorem is on the convergence rate under the spectral norm. It implies the convergence rates of eigenvalues and eigenvectors, which are essential in principle component analysis for example. The convergence rate under the spectral norm may also be important for classification, for example linear/quadratic discriminant analysis as we illustrate in Section 5.

**Theorem 1.** *Let* $\lambda_{ni} = C_0\sqrt{\log p/n}$ *with* $C_0$ *being a sufficiently large number. Under (C1), and (C2) (or (C2*)), we have*

$$\left\|\hat{\mathbf{\Omega}} - \mathbf{\Omega}\right\|_2 \le CM_p s_p\sqrt{\frac{\log p}{n}}$$

*with probability greater than* $1 - O\left(p^{-1}\right)$ *(or* $1 - O(p^{-1} + n^{-\delta/8})$ *under (C2*)), where* $C > 0$ *depends only on* $c_0$, $\eta$, $C_0$ *and* $K$ *(or* $c_0$, $c_1, \gamma, \delta, C_0$ *and* $K$ *under (C2*)).*

**Remark 1.** If $M_p s_p n^{-1/2}\log^{1/2} p = o(1)$, then $\hat{\mathbf{\Omega}}$ is positive definite with probability tending to one. We can also revise $\hat{\mathbf{\Omega}}$ to $\hat{\mathbf{\Omega}}_\tau$ with

$$\hat{\mathbf{\Omega}}_\tau = \hat{\mathbf{\Omega}} + \tau\boldsymbol{I},$$

where $\tau = (|\Lambda_{\min}(\hat{\mathbf{\Omega}})| + n^{-1/2})1\{\Lambda_{\min}(\hat{\mathbf{\Omega}}) \leq 0\}$. By Theorem 1, assuming $\tau \leq CM_p s_p n^{-1/2} \log^{1/2} p$, we have with probability greater than $1 - O(p^{-1})$ (or $1 - O(p^{-1} + n^{-\delta/8})$) that

$$\left\| \hat{\mathbf{\Omega}}_\tau - \mathbf{\Omega} \right\|_2 \leq CM_p s_p \sqrt{\frac{\log p}{n}}.$$

Such a simple perturbation will make the revised estimator $\hat{\mathbf{\Omega}}_\tau$ to have a larger minimal eigenvalue, for stability concerns. The later results on support recovery and other norms will also hold under such a small perturbation.

**Remark 2.** [6] imposed the following irrepresentable condition on glasso: for some $0 < \alpha < 1$,

$$\left\| \mathbf{\Gamma}_{\Psi^c \Psi} \left( \mathbf{\Gamma}_{\Psi\Psi} \right)^{-1} \right\|_\infty \leq 1 - \alpha, \tag{5}$$

where $\Psi$ is the support of $\mathbf{\Omega}$, $\mathbf{\Gamma} = \mathbf{\Sigma} \otimes \mathbf{\Sigma}$, and $\otimes$ denotes the Kronecker matrix product. To make things concrete, we now compare our conditions using the examples given in [6]:

1. In the diamond graph, let $p = 4$, $\sigma_{ii} = 1$, $\sigma_{23} = 0$, $\sigma_{14} = 2\rho^2$ and $\sigma_{ij} = \rho$ for all $i \neq j$, $(i, j) \neq (2, 3)$ and $(2, 4)$. For this matrix, (5) is reduced to $4|\rho|(|\rho| + 1) < 1$ and so it requires $\rho \in (-0.208, 0.208)$. Our relaxed condition (4) only needs $\rho \in (-0.5, 0.5)$.
2. In the star graph, let $p = 4$, $\sigma_{ii} = 1$, $\sigma_{1,j} = \rho$ for $j = 2, 3, 4$, $\sigma_{ij} = \rho^2$ for $1 < i < j \leq 4$. For this model, (5) requires $|\rho|(|\rho| + 2) < 1$ (i.e. $\rho \in (-0.4142, 0.4142)$), while our condition (4) holds for all $\rho \in (-1, 1)$.

We have the following result on the convergence rates under the element-wise $l_\infty$ norm and the Frobenius norm.

**Theorem 2.** *Under the conditions of Theorem 1, we have with probability greater than $1 - O(p^{-1})$ under (C2) (or $1 - O(p^{-1} + n^{-\delta/8})$ under (C2*))*

$$\left| \hat{\mathbf{\Omega}} - \mathbf{\Omega} \right|_\infty \leq CM_p \sqrt{\frac{\log p}{n}} \tag{6}$$

*and*

$$\frac{1}{p} \left\| \hat{\mathbf{\Omega}} - \mathbf{\Omega} \right\|_F^2 \leq Cs_p \frac{\log p}{n}. \tag{7}$$

**Remark 3.** The convergence rate under the Frobenius norm does not depend on $M_p$. In comparison, [17] obtained the minimax lower bound, when $\boldsymbol{X} \sim N(\boldsymbol{\mu}, \boldsymbol{\Sigma})$,

$$\frac{1}{p} \min_{\hat{\boldsymbol{\Omega}}} \max_{\boldsymbol{\Omega} \in \mathcal{U}} \mathsf{E} \left\| \hat{\boldsymbol{\Omega}} - \boldsymbol{\Omega} \right\|_F^2 \geq C M_p^2 s_p \frac{\log p}{n}. \tag{8}$$

They also showed that this rate is achieved by sequentially running two CLIME estimators, where the second CLIME estimator uses the first CLIME estimate as input. Though CLIME allows a weaker sparsity condition where our $\ell_0$ ball bound $s_p$ in $\mathcal{U}$ is replaced by an $\ell_q$ ball bound $(0 \leq q < 1)$, our rate in (7) is faster than CLIME, because $M_p^2$ in (8) could grow with $p$. The faster rate is due to the fact that we consider the condition (4). Under a slightly stronger condition (5) (see Remark 2), [6] proved that the glasso estimator $\hat{\boldsymbol{\Omega}}_{glasso}$ has the following convergence rate

$$\frac{1}{p} \left\| \hat{\boldsymbol{\Omega}}_{glasso} - \boldsymbol{\Omega} \right\|_F^2 = O_P \left( \kappa_\Gamma^2 s_p \frac{\log p}{n} \right), \tag{9}$$

where $\kappa_\Gamma = \left\| (\boldsymbol{\Gamma}_{\Psi\Psi})^{-1} \right\|_{L_1}$. Our convergence rate is also faster than theirs in (9) if $\kappa_\Gamma \to \infty$.

*3.3. Support recovery*

As discussed in the introduction, support recovery is related to Gaussian graphical models. The support of $\boldsymbol{\Omega}$ is recovered by SCIO, with high probability by the following theorem. Recall $\Psi = \{(i,j) : \omega_{ij} \neq 0\}$ be the support of $\boldsymbol{\Omega}$, and similarly

$$\hat{\Psi} = \{(i,j) : \hat{\omega}_{ij} \neq 0\}.$$

The next theorem gives the result on support recovery.

**Theorem 3.** *(i). Under the conditions of Theorem 1, we have $\hat{\Psi} \subseteq \Psi$ with probability greater than $1 - O(p^{-1})$ under (C2) (or $1 - O(p^{-1} + n^{-\delta/8})$ under (C2*)). (ii). In addition, suppose that for a sufficiently large number $C > 0$,*

$$\min_{(i,j) \in \Psi} |\omega_{ij}| \geq C M_p \sqrt{\frac{\log p}{n}}. \tag{10}$$

*Then under the conditions of Theorem 1, we have $\hat{\Psi} = \Psi$ with probability greater than $1 - O(p^{-1})$ under (C2) (or $1 - O(p^{-1} + n^{-\delta/8})$ under (C2*)).*

The condition (10) on the signal strength is standard for support recovery, see [6], [13] for example. We also note that the CLIME method [13] requires an additional thresholding step for support recovery, while SCIO does not need this step.

## 4. Theory for data-driven penalty

This section analyzes a cross validation scheme for choosing the tuning parameter $\lambda_{ni}$, and we establish the theoretical justification of this data-driven procedure.

We consider the following cross validation method for simplicity, similar to the one analyzed in [14]. Divide the sample $\{\boldsymbol{X}_k; 1 \leq k \leq n\}$ into two subsamples at random. Let $n_1$ and $n_2 = n - n_1$ be the two sample sizes of the random splits satisfying $n_1 \asymp n_2 \asymp n$, and let $\hat{\boldsymbol{\Sigma}}_1^l$, $\hat{\boldsymbol{\Sigma}}_2^l$ be the sample covariance matrices from the two samples $n_1$ and $n_2$ respectively in the $l$th split, for $l = 1, \ldots, H$, where $H$ is a fixed integer. For each $i$, let $\hat{\boldsymbol{\beta}}_i^l(\lambda)$ be the estimator minimizing the average out-of-sample SCIO loss, over $\lambda$,

$$\hat{R}_i(\lambda) = \frac{1}{H}\sum_{v=1}^{H}\left[\frac{1}{2}(\hat{\boldsymbol{\beta}}_i^l(\lambda))^T\hat{\boldsymbol{\Sigma}}_2^l\hat{\boldsymbol{\beta}}_i^l(\lambda) - \boldsymbol{e}_i^T\hat{\boldsymbol{\beta}}_i^l(\lambda)\right] \tag{11}$$

where $\hat{\boldsymbol{\beta}}_i^l(\lambda)$ is calculated from the $n_1$ samples with a tuning parameter $\lambda$ to be determined. For implementation purposes, instead of searching for continuous $\lambda$, we will divide the interval $[0, 4]$ by a grid $\lambda_0 < \lambda_1 < \cdots < \lambda_N$, where $\lambda_i = \frac{4i}{N}$. The number 4 comes from the CLIME constraint, see the supplementary material. The tuning parameter on the grid is chosen by, for each $i$,

$$\hat{\lambda}_i = \arg\min_{0 \leq j \leq N} \hat{R}_i(\lambda_j). \tag{12}$$

It is important to note that the size $N$ should be sufficiently large but not too large, see the first two conditions on $N$ in Theorem 4, and the convergence rate will then hold even if we only perform cross validation on a grid. The choice of $\hat{\lambda}_i$ could be different for estimating each column of the precision matrix using the column loss function (11). This allows the procedure to adapt to the magnitude and sparsity of each column, compared with the standard glasso estimator with a single choice of $\lambda$ for the whole matrix. Though it is possible to specify different $\lambda$ for each column (even each entry) in glasso, searching over all possible combinations of $\lambda$'s over high dimensional grids, using a non-column-wise loss (e.g. the likelihood), is computationally untrackable. Our column loss thus provides a simple and computationally trackable alternative for choosing adaptive $\lambda$.

As described before, the complexity of the likelihood function may make it difficult to analyze the glasso estimator using cross validation. Though CLIME uses a constrained approach for estimation, its constrained objective function cannot be directly used for cross validation. [13] proposed to use the likelihood function as the cross validation loss, which makes it difficult to establish the theory of cross validated CLIME. For a different setting of estimating the covariance matrix, [14] obtained the convergence rate under the Frobenius norm, using covariance thresholding. The threshold is also based on sample splitting like ours. However, to the best of our knowledge, it has been an open problem on establishing the theoretical justification of cross validation when estimating the precision matrix. Theorem 4 below fills the gap, showing that the estimator based on $\hat{\lambda}_i$ from (12) attains the optimal rate under the Frobenius norm. For simplicity, we set $H = 1$ as in [14].

Our theory adopts the following condition on the sub-Gaussian distribution, which was use in [18] for example.

**(C3).** There exist positive numbers $\eta' > 0$ and $K' > 0$ such that

$$\max_{|\boldsymbol{v}|_2=1} \mathsf{E} \exp\left(\eta'(\boldsymbol{v}^T(\boldsymbol{X}-\boldsymbol{\mu}))^2\right) \le K'$$

This condition is slightly stronger than (C2), because our next theorem adapts to unknown $\boldsymbol{\Omega}$ using cross validation, instead of the theoretical choice $\lambda_{ni}$. It is easy to see that (C3) holds for the multivariate normal distribution as a special case.

Denote the unsymmetrized $\hat{\boldsymbol{\Omega}}_1^1 := (\hat{\omega}_{ij1}^1) = (\hat{\boldsymbol{\beta}}_1^1(\hat{\lambda}_1), \ldots, \hat{\boldsymbol{\beta}}_p^1(\hat{\lambda}_p))$ and recall the symmetrized matrix $\hat{\mathbf{\Omega}}^1$ as

$$\hat{\omega}_{ij}^1 = \hat{\omega}_{ji}^1 = \hat{\omega}_{ij1}^1 1\{|\hat{\omega}_{ij1}^1| < |\hat{\omega}_{ji1}^1|\} + \hat{\omega}_{ji1}^1 1\{|\hat{\omega}_{ij1}^1| \geq |\hat{\omega}_{ji1}^1|\}.$$

The following theorem shows that the estimator $\hat{\mathbf{\Omega}}^1 = (\hat{\omega}_{ij}^1)$ attains the minimax optimal rate under the Frobenius norm.

**Theorem 4.** *Under the conditions* $\log N = O(\log p)$, $\sqrt{n/\log p} = o(N)$, *and (C3), we have as* $n, p \to \infty$,

$$\frac{1}{p}\left\|\hat{\mathbf{\Omega}}^1 - \mathbf{\Omega}\right\|_F^2 = O_P\left(s_p \frac{\log p}{n}\right).$$

The convergence rate using cross validation is the same as (7) in Theorem 2 with the theoretical choice of $\lambda$. Using similar arguments in Theorem 4 of [14], this result can be extended to multiple folds $H > 1$. To the best of our knowledge, Theorem 4 is the first result on the theoretical justification of cross validation when estimating the sparse precision matrix.

## 5. Numerical examples

### 5.1. Algorithm

Recall that the SCIO estimator is obtained by applying symmetrization (2) to the solution from (1), where each column $\hat{\boldsymbol{\beta}}_i$ is given by the following

$$\hat{\boldsymbol{\beta}}_i = \underset{\boldsymbol{\beta}_i \in I\!R^p}{\arg\min} \left\{ \frac{1}{2}\boldsymbol{\beta}_i^T \hat{\mathbf{\Sigma}} \boldsymbol{\beta}_i - \boldsymbol{e}_i^T \boldsymbol{\beta}_i + \lambda \left|\boldsymbol{\beta}_i\right|_1 \right\} \tag{13}$$

for any $\lambda > 0$. We propose to employ an iterative coordinate descent algorithm to solve (13) for each $i$. In contrast, the R package glasso employs an outside loop over the columns of the precision matrix, while having another inside loop over the coordinates of each column. Our algorithm does not need an outside loop because our loss function is column-wise.

The iterative coordinate descent algorithm for each $i$ goes as follows. In each iteration, we fix all but one coordinate in $\boldsymbol{\beta}$, and optimize over that fixed coordinate. Without loss of generality, we consider optimizing over the $p$th coordinate $\beta_p$ while all other coordinates of $\boldsymbol{\beta}$ (denoted by $\boldsymbol{\beta}_{-p}$) are fixed. The solution is in an explicit form by the following simple proposition. The solution when fixing other coordinates is similar, simply by permuting the matrix. We then loop through the coordinates until the updates are smaller than a user-specified threshold, say $10^{-4}$.

**Proposition 1.** *Let the subvector partition* $\boldsymbol{\beta} = (\boldsymbol{\beta}_{-p}, \beta_p)$ *and partition* $\hat{\boldsymbol{\Sigma}}$ *accordingly as follows*

$$\hat{\boldsymbol{\Sigma}} = \begin{pmatrix} \hat{\boldsymbol{\Sigma}}_{11} & \hat{\boldsymbol{\Sigma}}_{12}^T \\ \hat{\boldsymbol{\Sigma}}_{12} & \hat{\Sigma}_{22} \end{pmatrix}.$$

*Fixing* $\boldsymbol{\beta}_{-p}$*, the minimizer of* (13) *is*

$$\beta_p = \mathcal{T}\left(1\{p = i\} - \boldsymbol{\beta}_{-p}^T \hat{\boldsymbol{\Sigma}}_{12}, \lambda\right) / \hat{\Sigma}_{22}$$

*where the soft thresholding rule* $\mathcal{T}(x, \lambda) = \operatorname{sign}(x) \max(|x| - \lambda, 0)$.

We implement this algorithm in an R package, *scio*, available on CRAN. All the following computation is performed using R on an AMD Opteron processor (2.6 GHz) with 32 Gb memory. The glasso and CLIME estimators are computed using its R packages *glasso* (version 1.7) and *clime* (version 0.4.1) respectively. The path-following strategy with warm-starts [11] is enabled in all methods.

### *5.2. Simulations*

In this section, we compare the performance with glasso and CLIME on several measures using simulated data. In order to compare the adaptivity of the procedures, the covariance matrices that generate the data all contain two block diagonals of different magnitude, where the second block is 4 times the first one. Similar examples were used in [15] in comparing adaptive covariance estimation. The first block is generated from the following models respectively.

1. **decay**: $\omega_{ij} = 0.6^{|i-j|}$.

2. **sparse**: Let the prototype $\mathbf{\Omega}_0 = \boldsymbol{O} + \delta \boldsymbol{I}$, where each off-diagonal entry in $\boldsymbol{O}$ is generated independently, and equals to 0.5 with probability 0.1 and 0 with probability 0.9. $\delta$ is chosen such that the conditional number (the ratio of maximal and minimal singular values of a matrix) equals to $p$. Finally, the block matrix is standardized to have unit diagonals.
3. **block**: A block diagonal matrix with block size 5 where each block has off-diagonal entries equal to 0.5 and diagonal 1. The resulting matrix is then randomly permuted.

100 independent and identically distributed observations constituting a training data set are generated from each multivariate Gaussian covariance model with mean zero, and 100 additional observations are generated from the same model as a validating data set. Using the training data alone, a series of penalized estimators with 50 different tuning parameters $\lambda$ is computed. For a fair comparison, we first pick the tuning parameters in glasso, CLIME, and SCIO to produce the smallest Bregman loss on the validation sample. The Bregman loss is defined by

$$L(\mathbf{\Sigma}, \mathbf{\Omega}) = \langle \mathbf{\Omega}, \mathbf{\Sigma} \rangle - \log \det(\mathbf{\Omega}).$$

We also compare with our cross validation scheme in Section 4, where the cross validation loss is the column-wise adaptive loss (11). The resulting estimator is denoted by SCIOcv. We consider different values of $p = 50, 100, 200, 400, 800, 1600$, and replicate 100 times.

Table 1 compares the estimation performance of SCIO, SCIOcv, CLIME, and glasso under the spectral norm and the Frobenius norm. It shows that SCIO and SCIOcv almost uniformly outperform all other methods under both norms. SCIO has better performance when $p \leq 400$, while SCIOcv has better performance when $p \geq 800$. The glasso estimator has the worst performance overall, but it has slightly improved performance than other methods in the block model for $p = 200$ and 400. The CLIME estimator has slightly worse performance than our estimators overall, except for a few cases.

As discussed before, support recovery carries important consequences for

Table 1: Comparison of average losses of SCIO, SCIOcv, CLIME, and glasso over 100 simulation runs. The best performance is highlighted in bold. All standard errors of the results are smaller than 0.1.

Spectral Norm

| | Decay | | | | Sparse | | | | Block | | | |
|---|---|---|---|---|---|---|---|---|---|---|---|---|
| p | SCIO | SCIOcv | CLIME | glasso | SCIO | SCIOcv | CLIME | glasso | SCIO | SCIOcv | CLIME | glasso |
| 50 | **10.00** | 11.24 | 11.62 | 12.10 | **2.73** | 4.03 | 5.70 | 3.86 | **7.24** | 9.55 | 8.03 | 9.61 |
| 100 | **11.89** | 12.68 | 12.29 | 13.11 | **4.51** | 5.57 | 6.54 | 5.70 | **9.63** | 9.78 | 9.13 | 9.77 |
| 200 | **12.88** | 13.46 | 12.91 | 13.84 | **7.93** | 8.31 | 8.43 | 8.48 | 9.88 | 9.85 | 10.05 | **9.83** |
| 400 | **13.63** | 13.87 | 14.09 | 14.07 | **10.88** | 11.60 | 11.63 | 11.11 | 9.92 | 9.91 | 10.31 | **9.87** |
| 800 | 14.13 | **14.05** | 14.10 | 14.71 | 15.58 | **15.48** | 15.60 | 16.08 | 9.96 | **9.95** | 10.01 | 10.63 |
| 1600 | 14.15 | **14.12** | **14.12** | 14.83 | 20.94 | **20.90** | 20.94 | 21.61 | 9.97 | **9.96** | 10.15 | 10.68 |

Frobenius Norm

| | Decay | | | | Sparse | | | | Block | | | |
|---|---|---|---|---|---|---|---|---|---|---|---|---|
| p | SCIO | SCIOcv | CLIME | glasso | SCIO | SCIOcv | CLIME | glasso | SCIO | SCIOcv | CLIME | glasso |
| 50 | **16.22** | 18.54 | 19.25 | 20.18 | **6.71** | 7.95 | 12.66 | 8.14 | **16.10** | 20.98 | 17.58 | 21.68 |
| 100 | **27.48** | 29.58 | 28.40 | 30.92 | **12.93** | 14.84 | 18.48 | 14.91 | **30.83** | 31.02 | 28.72 | 31.15 |
| 200 | 42.93 | 45.12 | **42.80** | 47.00 | **24.34** | 24.67 | 26.60 | 26.11 | 44.49 | 44.23 | 44.92 | **44.19** |
| 400 | **65.61** | 66.60 | 68.65 | 68.10 | **36.65** | 38.99 | 40.67 | 37.76 | 62.91 | 62.73 | 65.38 | **62.54** |
| 800 | 97.52 | **96.09** | 97.25 | 102.67 | 59.08 | **57.55** | 59.97 | 66.30 | 88.98 | **88.78** | 88.63 | 96.42 |
| 1600 | 138.09 | **136.90** | 137.74 | 147.11 | 83.85 | **82.87** | 84.50 | 96.90 | 125.85 | 125.64 | **125.41** | 137.27 |

graphical model estimation. The frequencies of correct zero/nonzero identification are summarized in Table 1 of the supplementary material. In there, the SCIO and SCIOcv estimates are sparser than the CLIME and glasso estimates in general. To further illustrate this, we plot the heatmaps of support recovery in Figure 1 using $p = 100$ as a representing example. These heatmaps confirm that the SCIO estimates usually contain less zeros than glasso and CLIME. By visual inspection, these SCIO estimates also tend to be closer to the truth, especially under the sparse model. In particular, they adapt to different magnitude. In contrast, glasso yields some interference patterns and artificial stripes, especially under the sparse model.

### *5.3. A genetic dataset on HIV-1 associated neurocognitive disorders*

Antiretroviral therapy (ART) has greatly reduced mortality and morbidity of HIV patients; however, HIV-1 associated neurocognitive disorders (HAND) are becoming common, which cause greatly degradation of life quality. We here apply our graphical models to a gene expression dataset [19] to study how their genetic interactions/pathways are altered between treated and untreated HAND patients, and compare with other methods using classification. The supplementary material includes the full description of the dataset, the modeling approach, and additional results.

Figure 3a compares classification accuracy between treated and untreated HAND. The results comparing HAND and controls are not shown because all methods have a constant area-under-the-curve value 1. Because the number of nonzero off-diagonal elements may depend on the different scales of the penalization parameters in each method, we plot the classification accuracy against the average percentages of nonzero off-diagonals of these two classes (treated and untreated), i.e. the average percentages of connected edges in two recovered graphical models for the treated and untreated respectively. The SCIOcv estimators (not shown) only differs from SCIO on how to pick $\lambda$ in a data-driven way, and thus it has the identical performance as SCIO under the same $\lambda$. This figure shows that in most cases SCIO outperforms glasso and CLIME when

Figure 1: Heatmaps of support recovery over 100 simulation runs (black is 100/100, white is 0/100).

Decay Model

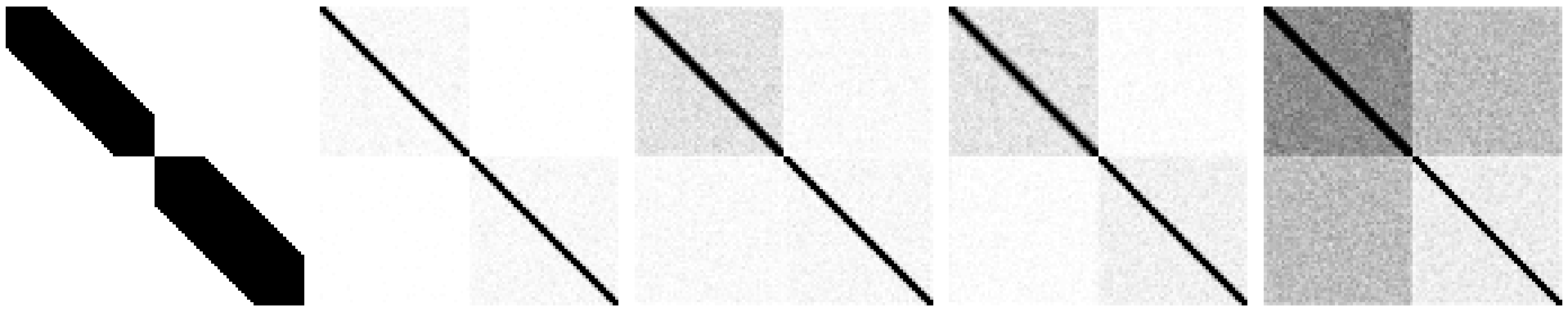

(a) Truth (b) SCIO (c) SCIOcv (d) CLIME (e) glasso

Sparse Model

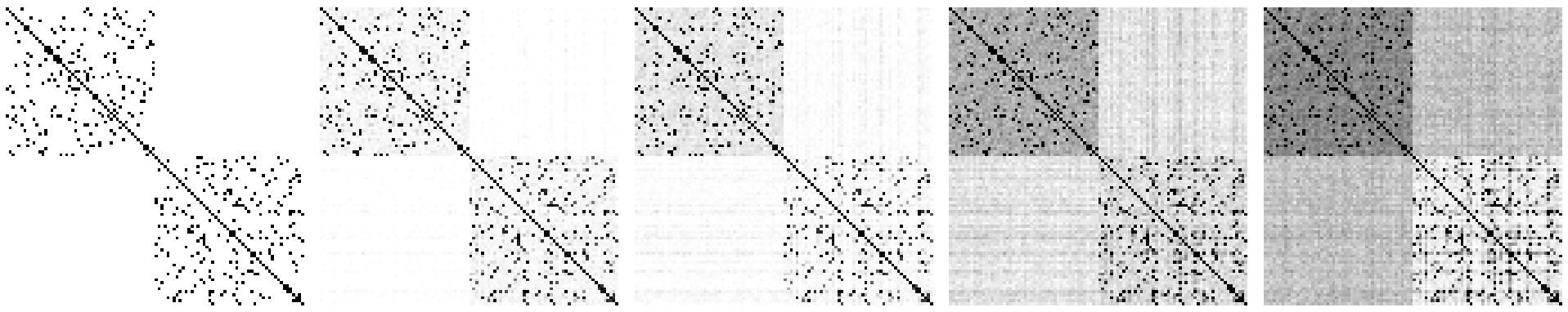

(f) Truth (g) SCIO (h) SCIOcv (i) CLIME (j) glasso

Block Model

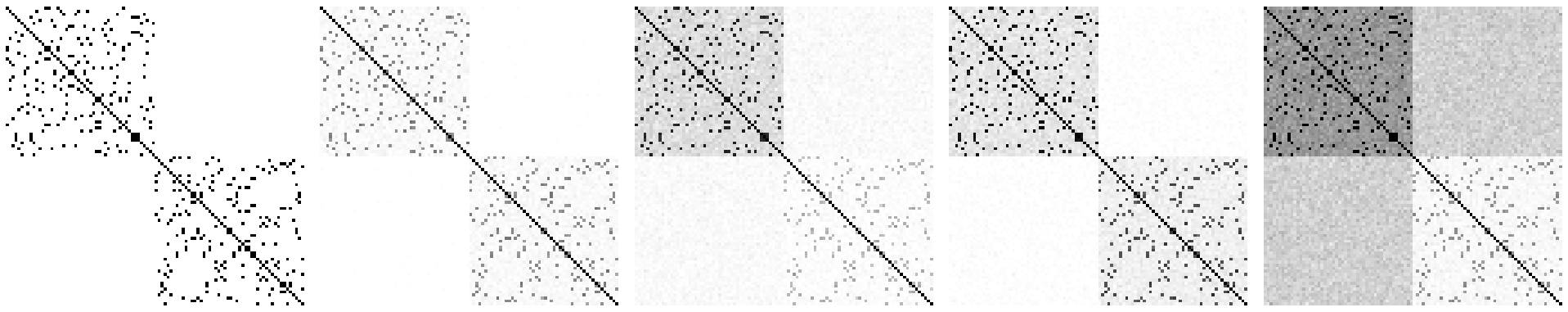

(k) Truth (l) SCIO (m) SCIOcv (n) CLIME (o) glasso

both methods use the same number of connected edges. The SCIO estimators are stable in classification performance even if the number of connected edges increases. We are not able to plot the performance of glasso with more than 14% connected edges (corresponding to small penalization parameters), because the glasso package does not converge within 120 hours. CLIME shows decreased performance when the number of connected edges increases. As a comparison with other classification algorithms, we use the same data to compare with a few other classification methods, including random forest [20], AIC penalized logistic regression, and $\ell_1$ penalized logistic regression with 5-fold cross validation. Their classification accuracies are 78.6%, 90.9% and 45.6% respectively. Our classification rule compares favorably with these competing methods on this dataset.

Figure 3b compares the running times against the percentages of connected edges. Because it is known that path-following algorithms may compute a sequence of solutions much faster than for a single one, we use 50 log-spaced penalization parameters from the largest (0% edges) to the designated percentages of edges, including 5%, 10%, 14%, 20%, 30%, 40%, 50% and 60%. As reported before, we are unable to plot the running times for glasso beyond 14% due to nonconvergence. SCIO takes about 2 seconds more than glasso when computing for 5% edges, but is much faster than glasso for 10% and more. For example, it compares favorably in the 14% case where SCIO takes only a quarter of the time of glasso. In general, the running time of SCIO grows linearly with the number of connected edges, while glasso shows exponential growth in computation time. CLIME is the slowest among all methods.

Figure 1 of the supplementary material compares the performance of support recovery, and it shows similar advantages of SCIO as in the simulations.

### *5.4. An fMRI dataset on attention deficit hyperactivity disorders*

Attention Deficit Hyperactivity Disorder (ADHD) causes substantial impairment among about 10% of school-age children in United States. A neuroimaging study showed that the correlations between brain regions are different between

Figure 2: Comparison of classification accuracy and running times using SCIO, CLIME and glasso for the HIV dataset. Red solid lines are SCIO, green dash lines are CLIME, and blue dotted lines are glasso.

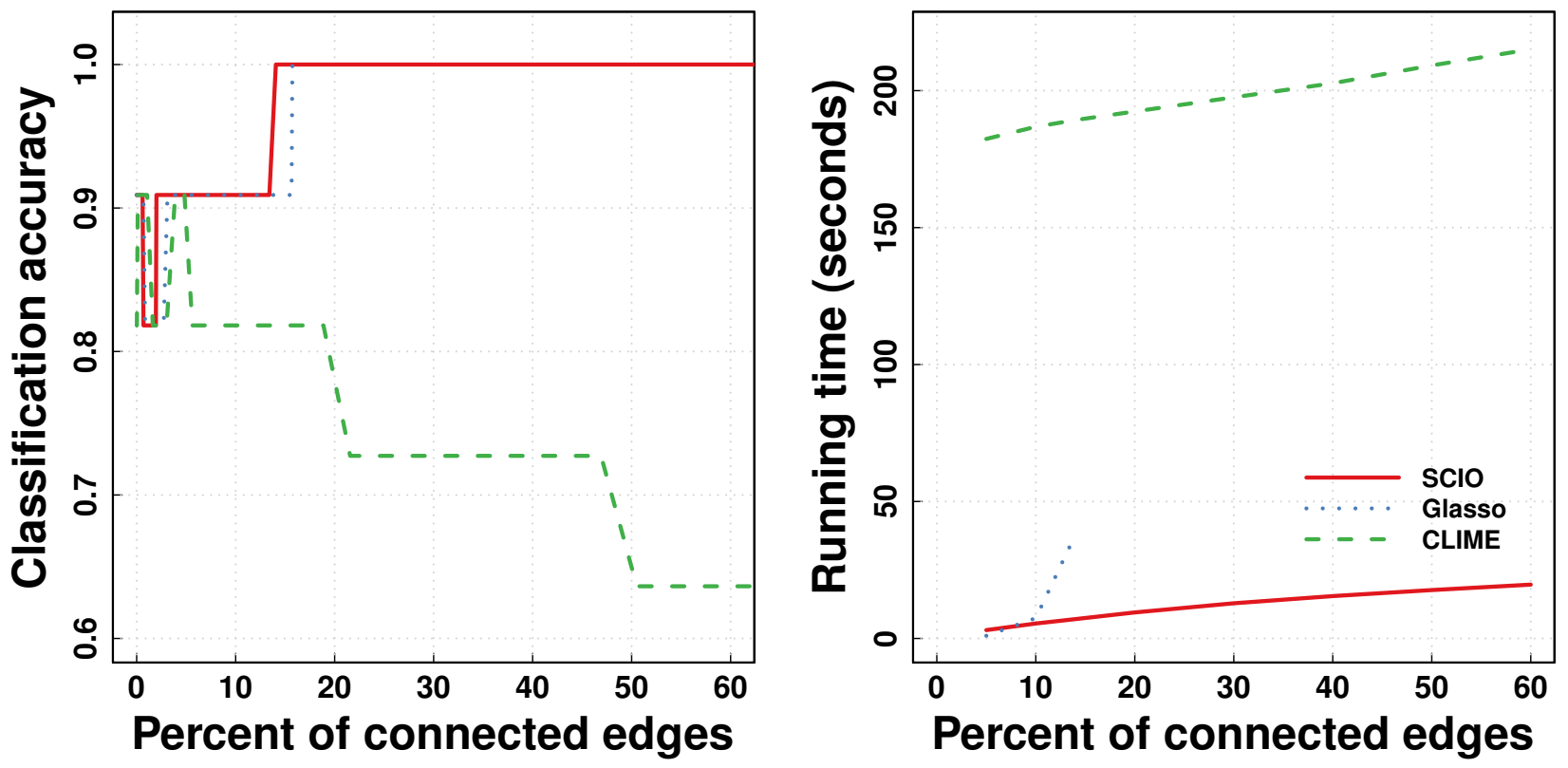

(a) Classification Accuracy

(b) Running times

typically developed children and children with such disorders [21]. The description of the data and additional results are provided in the supplementary material. In there, we compare the performance of support recovery using the data from each subject, and the results suggest that SCIO has competitive performance with CLIME and glasso in recovering brain connectivities for both healthy and ADHD children.

Figure 3 compares the running times of SCIO, CLIME, and glasso. Similar to the procedure described before, for each subject, we use path following algorithms in all methods up to the designated edge percentages, including 10%, 20%, 30%, 40%, 50% and 60%. This plot shows that the running times of SCIO grows almost linearly, and it is about 2 times faster than glasso with 60% connected edges. CLIME again is the slowest among all methods.

## 6. Discussion

It is possible to achieve adaptive estimation via other approaches. During the preparation of this paper, it comes to our attention that recently [22] applied a new adaptive penalized regression procedure, Scale Lasso, to the inverse covariance matrix estimation. [17] proposed an improved CLIME estimator, which runs the CLIME estimation sequentially twice. We instead analyzed cross validation as an alternative approach for this goal because cross validation remains to be popular among practitioners. It would be interesting to study the theory of cross validation for these other estimators, and to study if these adaptive approaches can also be applied to our loss.

Figure 3: Comparison of average (± 1 SE) running times for the ADHD dataset. The red solid line with circle marks is SCIO, the green dashed line with crosses is CLIME, and the blue dotted line with triangles is glasso.

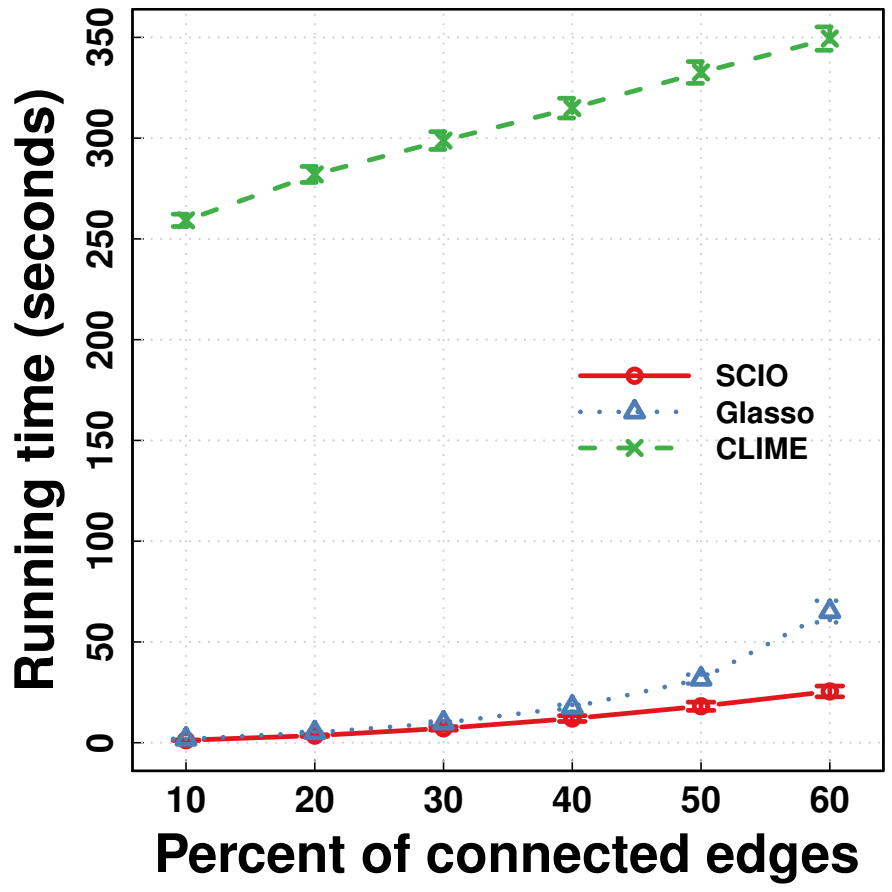

Choosing the tuning parameters is an important problem in the practice of penalization procedures, though most of the prior theoretical results are based on some theoretical assumptions of the tuning parameters. This paper is among the first to demonstrate that a cross validated estimator for the problem of precision matrix estimation achieves the $n^{-1/2}\log^{1/2} p$ rate under the Frobenius norm. This rate may not be improved in general, because it should be minimax optimal [17], though a rigorous justification is needed. We also note that the distribution condition (C3) in Theorem 4 is slightly stronger than (C2) and (C2*). It is an interesting problem to study if the result in Theorem 4 can be extended to more general distributions. Moreover, it would be interesting to study whether minimax rates can also be achieved under other matrix norms, such as the operator norm, using cross validation.

The rate for support recovery in Theorem 3 also coincides with the minimax optimal rate in [17]. However, $\mathcal{U}$ together with (4) is actually a smaller class than theirs. It would be interesting to explore if their minimax rate can be improved in this important sub-class. It would also be interesting to study if our results can be extended to their general matrix class.

We employ the $\ell_1$ norm to enforce sparsity due to computational concerns. It has been pointed out before that the $\ell_1$ penalty inheritably introduces biases, and thus it would be interesting to replace the $\ell_1$ norm by other penalty forms, such as Adaptive Lasso [23] or SCAD [9]. Such extensions should be easy to implement because our loss is column-wise, similar to penalized regression. We are currently implementing these variants for future releases of our R package.

There are several other interesting directions. It would be interesting to study the precision matrix estimation under the setting that the data are generated from statistical models, while the covariance estimation problem under this setting was studied by [24]. It is also of interest to consider extending SCIO to the nonparanormal family distributions [25].

Finally, this paper only considers the setting that all the data are observed. It is an interesting problem to study the inverse covariance matrix estimation when some observations are missing. It turns out that the SCIO procedure can also be applied to the missing data setting, with additional modifications. Due to the space limitation, we will report these results elsewhere.

## 7. Supplementary material

Supplementary material online includes the motivation of our estimator, additional descriptions of the numerical examples, and proof of the main results.

## 8. Acknowledgment

We would like to thank the Associate Editor and two anonymous referees for their very helpful comments, which have led to improved presentation of this paper. Weidong Liu was supported by the National Natural Science Foundation of China grants 11201298 and 11322107, the Program for Professor of Special Appointment (Eastern Scholar) at Shanghai Institutions of Higher Learning, Shanghai Pujiang Program, Foundation for the Author of National Excellent Doctoral Dissertation of China and a grant from Australian Research Council. Xi Luo was partially supported by the National Institutes of Health grants P01AA019072, P20GM103645, P30AI042853, R01NS052470, and S10OD016366, a Brown University Research Seed award, a Brown Institute for Brain Science Pilot award, a Brown University faculty start-up fund, and a developmental research award from Lifespan/Brown/Tufts Center for AIDS Research.

# Supplementary Material for "Fast and Adaptive Sparse Precision Matrix Estimation in High Dimensions"

Weidong Liu and Xi Luo*

September 21, 2014

This supplementary material provides the derivation of the method, additional numerical results, and proof of the main results in the main text.

# 1 Methodology

For simplicity, we start with a population covariance matrix $\mathbf{\Sigma}$, and we define a covariance loss function for every column $i = 1, 2, \dots, p$,

$$f_i(\mathbf{\Sigma}, \boldsymbol{B}) = \frac{1}{2}\boldsymbol{\beta}_i^T \mathbf{\Sigma} \boldsymbol{\beta}_i - \boldsymbol{e}_i^T \boldsymbol{\beta}_i, \tag{1}$$

where $\boldsymbol{B} = \left(\boldsymbol{\beta}_1, \boldsymbol{\beta}_2, \dots, \boldsymbol{\beta}_p\right)$, and each $\boldsymbol{\beta}_i$ is a column vector. Each function $f_i$ in (1) is strictly convex in $\boldsymbol{\beta}_i$ as $\mathbf{\Sigma}$ is strictly positive-definite; more importantly, the minimal values of each $f_i$ are achieved at $\boldsymbol{\beta}_i$'s that satisfy the following equality, for each $i$,

$$\mathbf{\Sigma}\boldsymbol{\beta}_i - \boldsymbol{e}_i = \mathbf{0}. \tag{2}$$

The quadratic function (1) is of the same form as the iterative conjugate gradient method that solves a linear system like (2). It is also straightforward to see that each column of the precision matrix $\mathbf{\Omega}$ satisfies these equalities, and thus minimizing all the loss functions in (1). In fact, $\mathbf{\Omega}$ is the unique solution of (2) if $\mathbf{\Sigma}$ is full rank, which can be seen by the inversion formula $\boldsymbol{\omega}_i = \mathbf{\Sigma}^{-1}\boldsymbol{e}_i = \mathbf{\Omega}\boldsymbol{e}_i$.

Certainly, the functions in (1) and the inversion formula cannot be directly applied to data, because the population covariance $\mathbf{\Sigma}$ is usually unknown. Thus, we replace with the sample covariance matrix $\hat{\mathbf{\Sigma}}$, to produce the sample based loss function of (1):

$$f_i(\hat{\mathbf{\Sigma}}, \boldsymbol{B}) = \frac{1}{2}\boldsymbol{\beta}_i^T \hat{\mathbf{\Sigma}} \boldsymbol{\beta}_i - \boldsymbol{e}_i^T \boldsymbol{\beta}_i.$$

One intuition is to minimize the above function, for every $i$, to produce an estimator for $\mathbf{\Omega}$. However, this is not possible because there may be multiple solutions if $\hat{\mathbf{\Sigma}}$ is not full rank. Moreover, it does not utilize the sparsity assumption of $\mathbf{\Omega}$. We will address these two issues momentarily.

Motivated by recent developments on using the $\ell_1$ norm in sparse precision matrix estimation (Friedman, Hastie, and Tibshirani, 2008; Cai, Liu and Luo, 2011), we add the $\ell_1$ penalty to the column loss function

$$\frac{1}{2}\boldsymbol{\beta}_i^T \hat{\mathbf{\Sigma}} \boldsymbol{\beta}_i - \boldsymbol{e}_i^T \boldsymbol{\beta}_i + \lambda_{ni} |\boldsymbol{\beta}_i|_1 \tag{3}$$

for $i = 1, 2, \ldots, p$, where the penalization parameter $\lambda_{ni} > 0$ and is allowed to be different from column to column, in order to adapt to the magnitude and sparsity of each column. Due to the shrinkage effect of the $\ell_1$ penalty, some coordinates of $\boldsymbol{\beta}_i$ may be shrunk to zero exactly. The loss function (3) is connected to the CLIME estimator (Cai, Liu and Luo, 2011). By taking the subgradient of (3), the minimal values satisfy the following constraint for $i = 1, 2, \ldots, p$,

$$\left| \hat{\mathbf{\Sigma}} \boldsymbol{\beta} - \boldsymbol{e}_i \right|_\infty \leq \lambda_{ni} \tag{4}$$

which is exactly the CLIME constraint.

# 2 Numerical examples

## 2.1 Simulations

Table 1 lists the frequencies of correct zero/nonzero identification by SCIO, SCIOcv, CLIME, and glasso, as discussed in the main text.

Table 1: Comparison of average support recovery (SD) of SCIO, SCIOcv, CLIME, and glasso over 100 simulation runs.

TN%

| | Decay | | | | Sparse | | | | Block | | | |
|---|---|---|---|---|---|---|---|---|---|---|---|---|
| p | SCIO | SCIOcv | CLIME | glasso | SCIO | SCIOcv | CLIME | glasso | SCIO | SCIOcv | CLIME | glasso |
| 50 | 98.57(0.72) | 97.22(0.84) | 88.84(2.40) | 76.18(3.06) | 85.16(1.62) | 97.73(0.54) | 64.60(7.20) | 83.16(2.45) | 80.60(1.93) | 95.67(0.98) | 86.69(2.16) | 87.40(5.12) |
| 100 | 99.71(0.13) | 98.97(0.21) | 96.55(0.68) | 86.03(1.60) | 91.40(0.44) | 98.73(0.20) | 79.51(3.89) | 86.69(1.16) | 97.34(2.44) | 98.69(0.32) | 95.97(0.65) | 96.72(1.27) |
| 200 | 99.98(0.02) | 99.61(0.06) | 99.40(0.08) | 94.97(2.20) | 96.11(0.29) | 99.42(0.07) | 93.70(1.49) | 90.55(0.66) | 99.97(0.12) | 99.71(0.07) | 99.67(0.86) | 99.03(0.36) |
| 400 | 100.00(0.00) | 99.84(0.02) | 100.0(0.00) | 98.90(0.16) | 98.66(0.06) | 99.72(0.03) | 98.62(0.27) | 95.60(0.44) | 100.00(0.01) | 99.94(0.01) | 99.98(0.01) | 99.68(0.11) |
| 800 | 100.00(0.00) | 99.94(0.01) | 100.0(0.00) | 95.29(0.06) | 100.00(0.00) | 99.86(0.01) | 100.0(0.00) | 94.20(0.38) | 100.00(0.00) | 99.98(0.00) | 100.0(0.00) | 95.85(0.06) |
| 1600 | 100.00(0.00) | 99.98(0.00) | 100.0(0.00) | 97.41(0.11) | 100.00(0.00) | 99.95(0.00) | 100.0(0.00) | 96.72(0.26) | 100.00(0.00) | 99.99(0.00) | 100.0(0.00) | 97.20(0.02) |

TP%

| | Decay | | | | Sparse | | | | Block | | | |
|---|---|---|---|---|---|---|---|---|---|---|---|---|
| p | SCIO | SCIOcv | CLIME | glasso | SCIO | SCIOcv | CLIME | glasso | SCIO | SCIOcv | CLIME | glasso |
| 50 | 24.19(2.24) | 21.60(1.65) | 37.21(2.91) | 35.92(2.32) | 98.71(1.22) | 93.27(2.75) | 99.88(0.39) | 96.00(2.28) | 95.18(2.83) | 58.26(5.12) | 98.50(1.17) | 62.45(6.20) |
| 100 | 12.67(0.52) | 13.77(0.76) | 21.54(1.37) | 26.44(1.37) | 77.73(2.12) | 75.73(2.50) | 97.05(1.06) | 83.55(2.66) | 31.09(10.94) | 41.94(3.33) | 85.56(3.07) | 48.98(3.48) |
| 200 | 10.14(0.26) | 9.92(0.38) | 12.76(0.32) | 16.15(3.46) | 41.20(1.68) | 29.78(1.33) | 62.99(3.58) | 62.98(1.73) | 20.02(0.11) | 30.11(1.70) | 39.17(2.27) | 38.81(3.11) |
| 400 | 7.14(0.78) | 7.84(0.18) | 3.46(0.00) | 8.81(0.37) | 10.68(0.39) | 12.03(0.44) | 24.17(2.17) | 33.83(1.41) | 20.00(0.01) | 24.63(0.75) | 23.70(0.72) | 32.15(2.02) |
| 800 | 3.40(0.00) | 6.81(0.08) | 3.40(0.00) | 15.43(0.24) | 2.44(0.00) | 5.02(0.16) | 2.41(0.01) | 25.50(0.79) | 20.00(0.00) | 22.47(0.36) | 19.99(0.02) | 52.19(0.56) |
| 1600 | 3.37(0.00) | 6.32(0.05) | 3.36(0.00) | 12.13(0.25) | 1.24(0.00) | 1.87(0.05) | 1.23(0.80) | 12.82(0.52) | 20.00(0.00) | 21.42(0.22) | 19.98(0.02) | 47.98(0.31) |

## 2.2 A genetic dataset on HIV-1 associated neurocognitive disorders

Borjabad et al (2011) analyzed gene expression arrays on post-mortem brain tissues. They showed that patients with HAND on antiretroviral therapy have many fewer and milder gene expression changes than untreated patients, and these genes are postulated to regulate several important genetic pathways. Their dataset is publicly available from Gene Expression Ominibus (GEO) under the serial number GSE28160. We here apply our method to study how their genetic interactions/pathways are altered between treated and untreated patients, and compare with other methods using classification, due to lack of the golden truth.

This dataset contains gene expression profiles of post-mortem brain tissues under two biological replications. The first replication contains 6 control (healthy) samples, 7 treated HAND samples, and 8 untreated HAND samples; the second replication contains 3 controls, 5 treated, and 6 untreated. The data are preprocessed by GEO and then log-transformed using Bioconductor in R. We will use the first replications as a training set, and test the performance of classifying 3 classes on the second replications. The class label is denoted by $q$, where $q = 1, 2, 3$ for control, treated and untreated respectively. The model building procedure is similar to Cai, Liu and Luo (2011). On the training data, we first compare pair-wise mean differences between 3 classes for each gene using Wilcoxon's tests, and select the top 100 genes with the most significant p-values in testing any pair of classes. Based on these 100 genes and the training data, we estimate the inverse covariance matrix $\hat{\boldsymbol{\Omega}}_q$ for each class $q$ using SCIO, CLIME, and glasso. To classify a new observation $X$ from the testing dataset, we employ a classification score for each pair of class $(q, q')$, which is defined as the log-likelihood difference (ignoring constant factors)

$$s_{q,q'}(\boldsymbol{X}) = -\left(\boldsymbol{X} - \overline{\boldsymbol{X}}_q\right)^T \hat{\boldsymbol{\Omega}}_q \left(\boldsymbol{X} - \overline{\boldsymbol{X}}_q\right) + \left(\boldsymbol{X} - \overline{\boldsymbol{X}}_{q'}\right)^T \hat{\boldsymbol{\Omega}}_{q'} \left(\boldsymbol{X} - \overline{\boldsymbol{X}}_{q'}\right) \\ + \log\det\left(\hat{\boldsymbol{\Omega}}_q\right) - \log\det\left(\hat{\boldsymbol{\Omega}}_{q'}\right)$$

where $\overline{\boldsymbol{X}}_j$ is the mean vector for class $j$ using the training data, $j = q, q'$ and $q \neq q'$. This score is essentially the logarithm of the likelihood ratios under two estimated multivariate normals. Because each class has almost the same number of observations in the training, we will assign the label $q$ if $s_{q,q'} > 0$ and $q'$ otherwise.

Figure 1 plots the support maps with a representing case of 10% connected edges using both SCIO, CLIME, and glasso. Each label has different connection patterns as shown by all these methods, and all methods share similar patterns by visual inspection. However, it should be noted that glasso tends to have stripes in the support, which is also observed in simulations.

### 2.3 An fMRI dataset on attention deficit hyperactivity disorders

The ADHD-200 project (http://fcon_1000.projects.nitrc.org/indi/adhd200/) released a resting-state fMRI dataset of healthy controls and ADHD children. We apply our method using the data in one of the participating center, Kennedy Krieger Institute. There are 61 typically-developing controls (HC), and 22 ADHD cases. The fMRI data were preprocessed by from neurobureau (http://www.nitrc.org/plugins/mwiki/index.php/neurobureau:AthenaPipeline), and the preprocessing steps are described on the same website. After preprocessing, we have 148 time points from each of 116 brain regions, for each subject. We will use the data of each subject to estimate the precision matrix. We choose the precision matrix instead of the covariance because it is more relevant to direct connections rather than indirect ones.

## 3 Proof of the main results

To prove the main results, we need the following lemmas. The first one comes from (28) and (33) in Cai, Liu and Luo (2011).

**Lemma 1** *Let $\mathbf{\Sigma} = (\sigma_{ij})_{p\times p}$ and the sample covariance $\hat{\mathbf{\Sigma}} = (\hat{\sigma}_{ij})_{p\times p}$. We have for some $C > 0$,*

$$P\Big(\max_{1\le i,j\le p}\{|\hat{\sigma}_{ij} - \sigma_{ij}|/(\sigma_{ii}^{1/2}\sigma_{jj}^{1/2})\} \ge C\sqrt{\frac{\log p}{n}}\Big) = O(p^{-1})$$

*under (C2), and*

$$P\Big(\max_{1\le i,j\le p}\{|\hat{\sigma}_{ij} - \sigma_{ij}|/(\sigma_{ii}^{1/2}\sigma_{jj}^{1/2})\} \ge C\sqrt{\frac{\log p}{n}}\Big) = O(p^{-1} + n^{-\delta/8})$$

*under (C2*).*

Figure 1: Comparison of support recovered by SCIO, CLIME, and glasso for the HIV dataset, when 10% of the edges are connected. Nonzeros are in black.

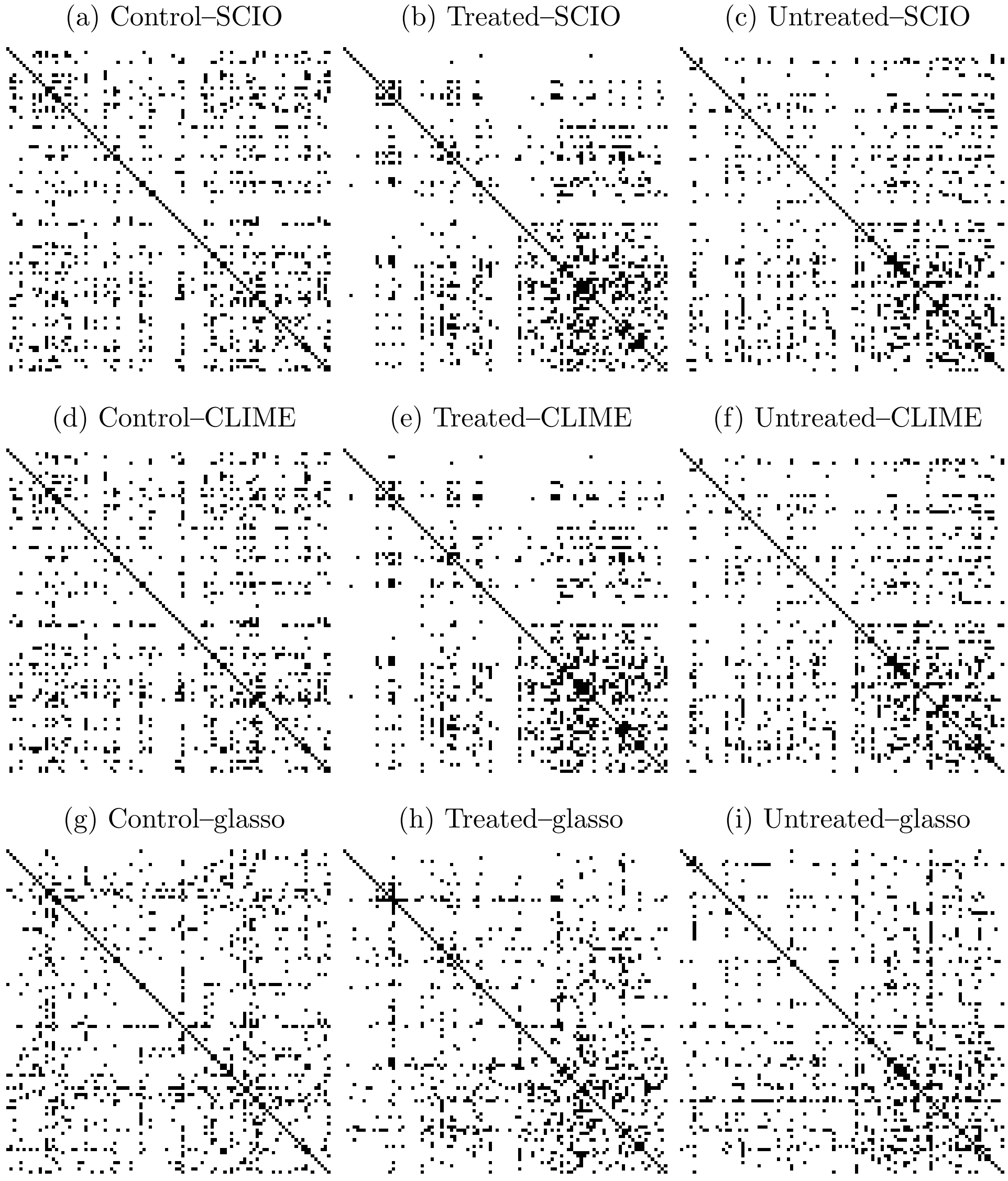

Figure 2: Comparison of support recovered by SCIO, CLIME, and glasso for the ADHD dataset, when 30% of the edges are connected. Black is nonzero over 100% of subjects, and white is 0%.

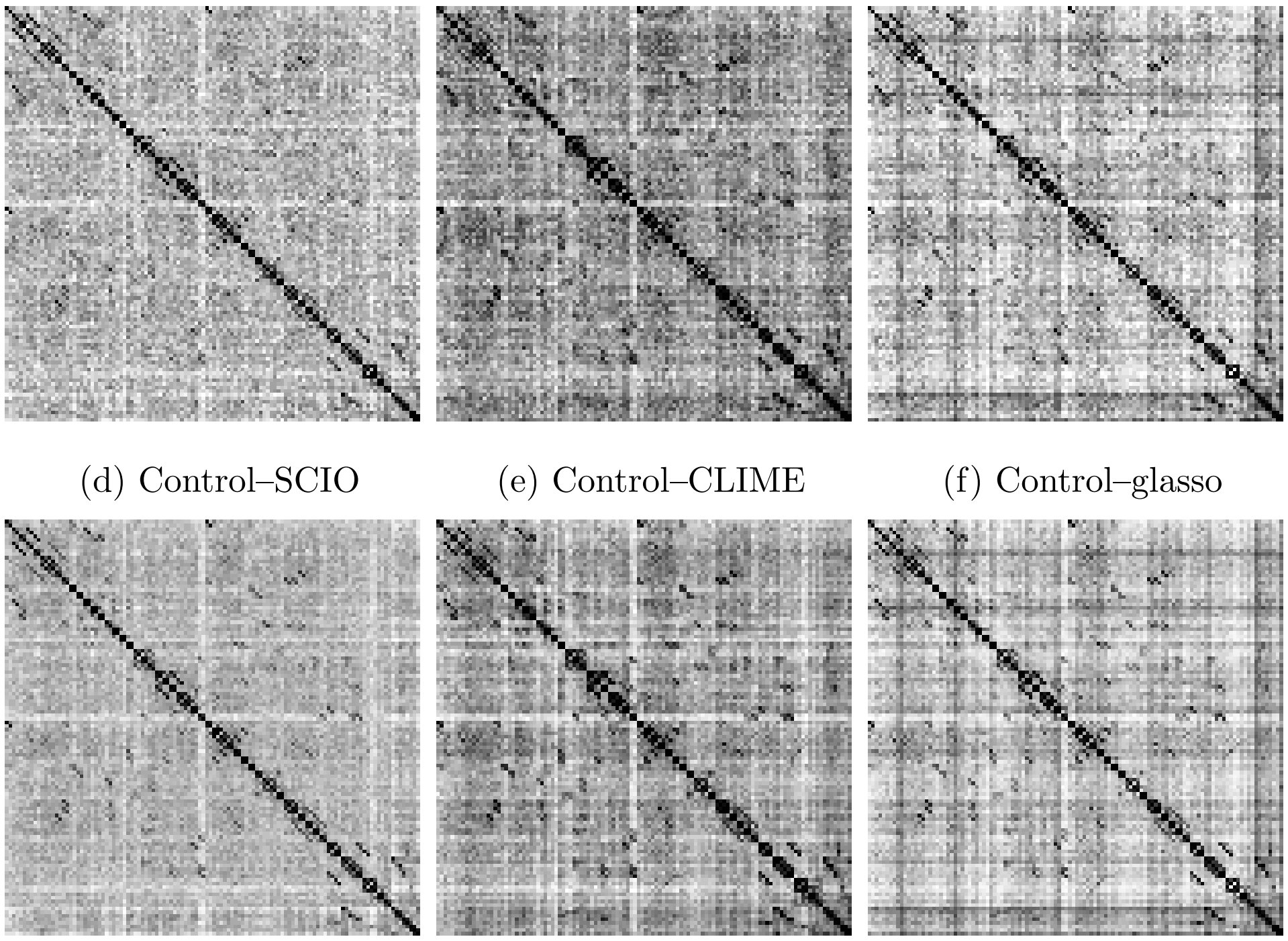

(a) ADHD–SCIO (b) ADHD–CLIME (c) ADHD–glasso

(d) Control–SCIO (e) Control–CLIME (f) Control–glasso

Let $\mathbf{\Omega} = (\omega_{ij}) = (\boldsymbol{\omega}_1, \ldots, \boldsymbol{\omega}_p)$, $\mathcal{S}_i$ be the support of $\boldsymbol{\omega}_i$ and $\boldsymbol{\omega}_{\mathcal{S}_i} = (\omega_{ji}; j \in \mathcal{S}_i)^T$. We will also need the following lemma from Cai, Liu and Zhou (2011).

**Lemma 2** *Assume $c_0^{-1} \leq \Lambda_{\min}(\mathbf{\Omega}) \leq \Lambda_{\max}(\mathbf{\Omega}) \leq c_0$. We have for some $C > 0$,*

$$P\Big(\max_{1\leq i\leq p} |\hat{\mathbf{\Sigma}}_{\mathcal{S}_i\times\mathcal{S}_i}\boldsymbol{\omega}_{\mathcal{S}_i} - \boldsymbol{e}_{\mathcal{S}_i}|_\infty \geq C\sqrt{\frac{\log p}{n}}\Big) = O(p^{-1})$$

*if (C2) holds;*

$$P\Big(\max_{1\leq i\leq p} |\hat{\mathbf{\Sigma}}_{\mathcal{S}_i\times\mathcal{S}_i}\boldsymbol{\omega}_{\mathcal{S}_i} - \boldsymbol{e}_{\mathcal{S}_i}|_\infty \geq C\sqrt{\frac{\log p}{n}}\Big) = O(p^{-1} + n^{-\delta/8})$$

*if (C2*) holds.*

**Proof of Theorem 1.** For the solution $\hat{\boldsymbol{\beta}}_i$, it satisfies that

$$\hat{\mathbf{\Sigma}}\hat{\boldsymbol{\beta}}_i - \boldsymbol{e}_i = -\lambda_{ni}\hat{\boldsymbol{Z}}_i,$$

where $\hat{\boldsymbol{Z}}_i =: (\hat{Z}_{1i}, \ldots, \hat{Z}_{pi})^T$ is the subdifferential $\partial|\hat{\boldsymbol{\beta}}_i|_1$ satisfying

$$\hat{Z}_{ji} = \begin{cases} 1, & \hat{\beta}_{ji} > 0; \\ -1, & \hat{\beta}_{ji} < 0; \\ \in [-1, 1], & \hat{\beta}_{ji} = 0. \end{cases}$$

Define $\hat{\boldsymbol{\beta}}_i^o$ be the solution of the following optimization problem:

$$\hat{\boldsymbol{\beta}}_i^o = \mathop{\arg\min}_{\mathbf{supp}(\boldsymbol{\beta})\subseteq\mathcal{S}_i} \Big\{\frac{1}{2}\boldsymbol{\beta}^T\hat{\mathbf{\Sigma}}\boldsymbol{\beta} - \boldsymbol{e}_i^T\boldsymbol{\beta} + \lambda_{ni}|\boldsymbol{\beta}|_1\Big\},$$

where $\mathbf{supp}(\boldsymbol{\beta})$ denotes the support of $\boldsymbol{\beta}$. We will show that $\hat{\boldsymbol{\beta}}_i = \hat{\boldsymbol{\beta}}_i^o$ with high probability.

Let $\hat{\boldsymbol{Z}}_{\mathcal{S}_i}^o$ is the subdifferential $\partial|\hat{\boldsymbol{\beta}}_i^o|_1$ on $\mathcal{S}_i$. We define the vector $\tilde{\boldsymbol{Z}}_i = (\tilde{Z}_{1i}, \ldots, \tilde{Z}_{pi})^T$ by letting $\tilde{Z}_{ji} = \hat{Z}_{ji}^o$ for $j \in \mathcal{S}_i$ and

$$\tilde{Z}_{ji} = -\lambda_{ni}^{-1}(\hat{\mathbf{\Sigma}}\hat{\boldsymbol{\beta}}_i^o)_j \quad \text{for } j \in \mathcal{S}_i^c.$$

By Lemma 3 proved momentarily, for $j \in \mathcal{S}_i^c$ and some $r < 1$,

$$|\tilde{Z}_{ji}| \leq r < 1 \tag{5}$$

with probability greater than $1 - O(p^{-1})$ under (C2) (or $1 - O(p^{-1} + n^{-\delta/8})$ under (C2*)). By this primal-dual witness construction and (9), the theorem is proved. ▮

The following lemma is employed when proving Theorem 1.

**Lemma 3** *With probability greater than* $1 - O(p^{-1})$ *under (C2) (or* $1 - O(p^{-1} + n^{-\delta/8})$ *under (C2*)), we have*

$$|\tilde{Z}_{ji}| < 1 - \alpha/2$$

*uniformly for* $j \in \mathcal{S}_i^c$.

**Proof.** By the definition of $\tilde{\boldsymbol{Z}}_i$, we have

$$\hat{\boldsymbol{\Sigma}}_{\mathcal{S}_i\mathcal{S}_i}\hat{\boldsymbol{\beta}}^o_{\mathcal{S}_i} - \boldsymbol{e}_{\mathcal{S}_i} = -\lambda_{ni}\tilde{\boldsymbol{Z}}_{\mathcal{S}_i} \tag{6}$$

and

$$\hat{\boldsymbol{\Sigma}}_{\mathcal{S}_i^c\mathcal{S}_i}\hat{\boldsymbol{\beta}}^o_{\mathcal{S}_i} \quad = \quad -\lambda_{ni}\tilde{\boldsymbol{Z}}_{\mathcal{S}_i^c}. \tag{7}$$

Write (6) as

$$\boldsymbol{\Sigma}_{\mathcal{S}_i\mathcal{S}_i}(\hat{\boldsymbol{\beta}}^o_{\mathcal{S}_i} - \boldsymbol{\omega}_{\mathcal{S}_i}) + (\hat{\boldsymbol{\Sigma}}_{\mathcal{S}_i\mathcal{S}_i} - \boldsymbol{\Sigma}_{\mathcal{S}_i\mathcal{S}_i})(\hat{\boldsymbol{\beta}}^o_{\mathcal{S}_i} - \boldsymbol{\omega}_{\mathcal{S}_i}) + \hat{\boldsymbol{\Sigma}}_{\mathcal{S}_i\mathcal{S}_i}\boldsymbol{\omega}_{\mathcal{S}_i} - \boldsymbol{e}_{\mathcal{S}_i} = -\lambda_{ni}\tilde{\boldsymbol{Z}}_{\mathcal{S}_i}.$$

This implies that

$$\hat{\boldsymbol{\beta}}^o_{\mathcal{S}_i} - \boldsymbol{\omega}_{\mathcal{S}_i} = \boldsymbol{\Sigma}^{-1}_{\mathcal{S}_i\mathcal{S}_i}\Big( - \lambda_n\tilde{\boldsymbol{Z}}_{\mathcal{S}_i} - (\hat{\boldsymbol{\Sigma}}_{\mathcal{S}_i\mathcal{S}_i} - \boldsymbol{\Sigma}_{\mathcal{S}_i\mathcal{S}_i})(\hat{\boldsymbol{\beta}}^o_{\mathcal{S}_i} - \boldsymbol{\omega}_{\mathcal{S}_i}) - \hat{\boldsymbol{\Sigma}}_{\mathcal{S}_i\mathcal{S}_i}\boldsymbol{\omega}_{\mathcal{S}_i} + \boldsymbol{e}_{\mathcal{S}_i}\Big). \tag{8}$$

By (3) of the main text, Lemma 1 and Lemma 2, we have with probability greater than $1 - O(p^{-1})$ (or $1 - O(p^{-1} + n^{-\delta/8})$),

$$|\hat{\boldsymbol{\beta}}^o_{\mathcal{S}_i} - \boldsymbol{\omega}_{\mathcal{S}_i}|_2 \le C\sqrt{s_p \log p/n} + o(1)|\hat{\boldsymbol{\beta}}^o_{\mathcal{S}_i} - \boldsymbol{\omega}_{\mathcal{S}_i}|_2.$$

This implies that

$$|\hat{\boldsymbol{\beta}}^o_{\mathcal{S}_i} - \boldsymbol{\omega}_{\mathcal{S}_i}|_2 \le C\sqrt{s_p \log p/n}. \tag{9}$$

By (7) and the above equation, we have

$$
\begin{aligned}
-\tilde{\boldsymbol{Z}}_{\mathcal{S}_i^c} &= \frac{1}{\lambda_n}\hat{\boldsymbol{\Sigma}}_{\mathcal{S}_i^c\mathcal{S}_i}(\hat{\boldsymbol{\beta}}^o_{\mathcal{S}_i}-\boldsymbol{\omega}_{\mathcal{S}_i})+\frac{1}{\lambda_n}(\hat{\boldsymbol{\Sigma}}_{\mathcal{S}_i^c\mathcal{S}_i}-\boldsymbol{\Sigma}_{\mathcal{S}_i^c\mathcal{S}_i})\boldsymbol{\omega}_{\mathcal{S}_i} \\
&= \frac{1}{\lambda_n}(\hat{\boldsymbol{\Sigma}}_{\mathcal{S}_i^c\mathcal{S}_i}-\boldsymbol{\Sigma}_{\mathcal{S}_i^c\mathcal{S}_i})(\hat{\boldsymbol{\beta}}^o_{\mathcal{S}_i}-\boldsymbol{\omega}_{\mathcal{S}_i})-\boldsymbol{\Sigma}_{\mathcal{S}_i^c\mathcal{S}_i}\boldsymbol{\Sigma}^{-1}_{\mathcal{S}_i\mathcal{S}_i}\tilde{\boldsymbol{Z}}_{\mathcal{S}_i} \\
&\quad -\frac{1}{\lambda_n}\boldsymbol{\Sigma}_{\mathcal{S}_i^c\mathcal{S}_i}\boldsymbol{\Sigma}^{-1}_{\mathcal{S}_i\mathcal{S}_i}(\hat{\boldsymbol{\Sigma}}_{\mathcal{S}_i\mathcal{S}_i}-\boldsymbol{\Sigma}_{\mathcal{S}_i\mathcal{S}_i})(\hat{\boldsymbol{\beta}}^o_{\mathcal{S}_i}-\boldsymbol{\omega}_{\mathcal{S}_i}) \\
&\quad -\frac{1}{\lambda_n}\boldsymbol{\Sigma}_{\mathcal{S}_i^c\mathcal{S}_i}\boldsymbol{\Sigma}^{-1}_{\mathcal{S}_i\mathcal{S}_i}(\hat{\boldsymbol{\Sigma}}_{\mathcal{S}_i\mathcal{S}_i}\boldsymbol{\omega}_{\mathcal{S}_i}-\boldsymbol{e}_{\mathcal{S}_i}) \\
&\quad +\frac{1}{\lambda_n}(\hat{\boldsymbol{\Sigma}}_{\mathcal{S}_i^c\mathcal{S}_i}-\boldsymbol{\Sigma}_{\mathcal{S}_i^c\mathcal{S}_i})\boldsymbol{\omega}_{\mathcal{S}_i}.
\end{aligned}
$$

Since $\|\boldsymbol{\Sigma}_{\mathcal{S}_i^c\mathcal{S}_i}\boldsymbol{\Sigma}^{-1}_{\mathcal{S}_i\mathcal{S}_i}\|_\infty \le 1-\alpha$ and $|\tilde{\boldsymbol{Z}}_{\mathcal{S}_i}|_\infty \le 1$, we have $|\boldsymbol{\Sigma}_{\mathcal{S}_i^c\mathcal{S}_i}\boldsymbol{\Sigma}^{-1}_{\mathcal{S}_i\mathcal{S}_i}\tilde{\boldsymbol{Z}}_{\mathcal{S}_i}|_\infty \le 1-\alpha$. By (9) and Lemma 1, we obtain that with probability greater than $1-O(p^{-1})$ (or $1-O(p^{-1}+n^{-\delta/8})$)

$$
|(\hat{\boldsymbol{\Sigma}}_{\mathcal{S}_i^c\mathcal{S}_i}-\boldsymbol{\Sigma}_{\mathcal{S}_i^c\mathcal{S}_i})(\hat{\boldsymbol{\beta}}^o_{\mathcal{S}_i}-\boldsymbol{\omega}_{\mathcal{S}_i})|_\infty \le Cs_p\log p/n. \tag{10}
$$

This, together with Lemma 2, implies (5). ∎

**Proof of Theorems 2 and 3**. By the proof of Theorem 1, we have $\hat{\boldsymbol{\beta}}_i=\hat{\boldsymbol{\beta}}^o_i$. Reorganize terms to yield that

$$
\hat{\boldsymbol{\beta}}_i-\boldsymbol{\omega}_i=\boldsymbol{\Sigma}^{-1}\Big(-\lambda_n\hat{\boldsymbol{Z}}_i-(\hat{\boldsymbol{\Sigma}}-\boldsymbol{\Sigma})(\hat{\boldsymbol{\beta}}_i-\boldsymbol{\omega}_i)-\hat{\boldsymbol{\Sigma}}\boldsymbol{\omega}_i+\boldsymbol{e}_i\Big). \tag{11}
$$

By (9) and Lemma 1, we obtain that with probability greater than $1-O(p^{-1})$ (or $1-O(p^{-1}+n^{-\delta/8})$),

$$
|(\hat{\boldsymbol{\Sigma}}-\boldsymbol{\Sigma})(\hat{\boldsymbol{\beta}}_i-\boldsymbol{\omega}_i)|_\infty \le Cs_p\log p/n. \tag{12}
$$

Thus,

$$
|\hat{\boldsymbol{\beta}}_i-\boldsymbol{\omega}_i|_\infty \le CM_p\sqrt{\frac{\log p}{n}}.
$$

This proves (6). By (9) and the inequality $\|\hat{\boldsymbol{\Omega}}-\boldsymbol{\Omega}\|_F^2 \le 2\sum_{j=1}^p|\hat{\boldsymbol{\beta}}_i-\boldsymbol{\omega}_i|_2^2$, we obtain (7). Theorem 3 (i) follows from the proof of Theorem 1. Theorem 3 (ii) follows from Theorem 2 and the lower bound condition on $\min_{(i,j)\in\Psi}|\omega_{ij}|$. ∎

**Proof of Theorem 4**. Let

$$\hat{\boldsymbol{\beta}}_i = \underset{\boldsymbol{\beta}\in I\!R^p}{\arg\min}\left\{\frac{1}{2}\boldsymbol{\beta}^T\hat{\Sigma}_1^1\boldsymbol{\beta} - \boldsymbol{e}_i^T\boldsymbol{\beta} + \lambda_{ni}|\boldsymbol{\beta}|_1\right\}$$

with the theoretical $\lambda_{ni} = C\sqrt{\log p/n} \in \{\lambda_i, 1 \le i \le N\}$ and $C$ is sufficiently large. Then by the proofs of Theorem 1 and 2, we have with probability greater than $1 - O(p^{-1})$,

$$\max_{1\le i\le p} |\hat{\boldsymbol{\beta}}_i - \boldsymbol{\omega}_i|_2^2 \le Cs_p\frac{\log p}{n}.$$

By the definition of $\hat{\boldsymbol{\beta}}_i^1$ with the cross validated $\hat{\lambda}_i$, we have

$$\frac{1}{2}(\hat{\boldsymbol{\beta}}_i^1)^T\hat{\boldsymbol{\Sigma}}_2^1\hat{\boldsymbol{\beta}}_i^1 - \boldsymbol{e}_i^T\hat{\boldsymbol{\beta}}_i^1 \le \frac{1}{2}(\hat{\boldsymbol{\beta}}_i)^T\hat{\boldsymbol{\Sigma}}_2^1\hat{\boldsymbol{\beta}}_i - \boldsymbol{e}_i^T\hat{\boldsymbol{\beta}}_i.$$

Set $\boldsymbol{D}_i = \hat{\boldsymbol{\beta}}_i^1 - \boldsymbol{\omega}_i$ and $\boldsymbol{D}_i^o = \hat{\boldsymbol{\beta}}_i - \boldsymbol{\omega}_i$. This implies that

$$\begin{aligned}
&\langle(\hat{\boldsymbol{\Sigma}}_2^1 - \boldsymbol{\Sigma})\boldsymbol{D}_i, \boldsymbol{D}_i\rangle + \langle\boldsymbol{\Sigma}\boldsymbol{D}_i, \boldsymbol{D}_i\rangle + 2\langle\hat{\boldsymbol{\Sigma}}_2^1\boldsymbol{\omega}_i - \boldsymbol{e}_i, \hat{\boldsymbol{\beta}}_i^1 - \hat{\boldsymbol{\beta}}_i\rangle \\
&\le \langle(\hat{\boldsymbol{\Sigma}}_2^1 - \boldsymbol{\Sigma})\boldsymbol{D}_i^o, \boldsymbol{D}_i^o\rangle + \langle\boldsymbol{\Sigma}\boldsymbol{D}_i^o, \boldsymbol{D}_i^o\rangle.
\end{aligned}$$

Lemma 4 proved later yields that

$$|\langle(\hat{\boldsymbol{\Sigma}}_2^1 - \boldsymbol{\Sigma})\boldsymbol{D}_i, \boldsymbol{D}_i\rangle| = O_P(1)|\boldsymbol{D}_i|_2^2\sqrt{\frac{\log N}{n}}$$

and

$$\langle\hat{\boldsymbol{\Sigma}}_2^1\boldsymbol{\omega}_{\cdot i} - \boldsymbol{e}_i, \hat{\boldsymbol{\beta}}_i^1 - \hat{\boldsymbol{\beta}}_i\rangle = O_P(1)|\hat{\boldsymbol{\beta}}_i^1 - \hat{\boldsymbol{\beta}}_i|_2\sqrt{\frac{\log N}{n}}.$$

Thus,

$$|\boldsymbol{D}_i|_2^2 \le O_P\Big(\sqrt{\frac{\log N}{n}}\Big)(|\boldsymbol{D}_i|_2 + |\hat{\boldsymbol{\beta}}_i - \boldsymbol{\omega}_i|_2) + |\boldsymbol{D}_i^o|_2^2.$$

This proves the theorem. ∎

The following lemma is needed for proving Theorem 4.

**Lemma 4** *For any vector $\boldsymbol{v}$ with $|\boldsymbol{v}|_2 = 1$, we have*

$$\max_{1\le i\le N} |\langle(\hat{\boldsymbol{\Sigma}}_2^1 - \boldsymbol{\Sigma})\boldsymbol{v}, \boldsymbol{v}\rangle| = O_P\left(\sqrt{\frac{\log N}{n}}\right) \tag{13}$$

*and*

$$\max_{1\le i\le N} |\langle\hat{\boldsymbol{\Sigma}}_2^1 \boldsymbol{\omega}_i - \boldsymbol{e}_i, \boldsymbol{v}\rangle| = O_P\left(\sqrt{\frac{\log N}{n}}\right). \tag{14}$$

**Proof.** We will use the following identity

$$\begin{aligned}\langle(\hat{\boldsymbol{\Sigma}}_2^1 - \boldsymbol{\Sigma})\boldsymbol{v}, \boldsymbol{v}\rangle &= \langle(\boldsymbol{\Sigma}^{-1/2}\hat{\boldsymbol{\Sigma}}_2^1\boldsymbol{\Sigma}^{-1/2} - \boldsymbol{I})\boldsymbol{\Sigma}^{1/2}\boldsymbol{v}, \boldsymbol{\Sigma}^{1/2}\boldsymbol{v}\rangle \\ &= \langle(\boldsymbol{\Sigma}^{-1/2}\tilde{\boldsymbol{\Sigma}}_2^1\boldsymbol{\Sigma}^{-1/2} - \boldsymbol{I})\boldsymbol{\Sigma}^{1/2}\boldsymbol{v}, \boldsymbol{\Sigma}^{1/2}\boldsymbol{v}\rangle + (\boldsymbol{v}^T\bar{\boldsymbol{X}} - \boldsymbol{v}^T\boldsymbol{\mu})^2,\end{aligned}$$

where $\tilde{\boldsymbol{\Sigma}}_2^1 = \frac{1}{n_2}\sum_{k=1}^{n_2}(\boldsymbol{X}_k - \boldsymbol{\mu})(\boldsymbol{X}_k - \boldsymbol{\mu})^T$. We have

$$\langle(\tilde{\boldsymbol{\Sigma}}_2^1 - \boldsymbol{\Sigma})\boldsymbol{v}, \boldsymbol{v}\rangle = \frac{1}{n_2}\sum_{k=1}^{n_2}(\boldsymbol{v}^T(\boldsymbol{X}_k - \boldsymbol{\mu}))^2 - \boldsymbol{v}^T\boldsymbol{\Sigma}\boldsymbol{v}.$$

By (C3) and the exponential inequality in Lemma 1, for any $M > 0$, there exists some $C > 0$ such that

$$\max_{1\le i\le N} P\Big(\Big|\frac{1}{n_2}\sum_{k=1}^{n_2}(\boldsymbol{v}^T(\boldsymbol{X}_k - \boldsymbol{\mu}))^2 - \boldsymbol{v}^T\boldsymbol{\Sigma}\boldsymbol{v}\Big| \ge C\sqrt{\frac{\log N}{n}}\Big) = O(N^{-M}),$$

$$\max_{1\le i\le N} P\Big(|\boldsymbol{v}^T\bar{\boldsymbol{X}} - \boldsymbol{v}^T\boldsymbol{\mu}| \ge C\sqrt{\frac{\log N}{n}}\Big) = O(N^{-M}).$$

Hence, (13) is proved. (14) follows from the exponential inequality in Lemma 2. ∎

**Proof of Proposition 1.** The objective is equivalent to (after neglecting constant terms with respect to $\beta_p$)

$$\beta_p\boldsymbol{\beta}_{-p}^T\hat{\boldsymbol{\Sigma}}_{12} + \frac{1}{2}\beta_p^2\hat{\Sigma}_{22} - \beta_p 1\{p = i\} + \lambda|\beta_p|.$$

The minimizer then should have a subgradient equal to zero,

$$\boldsymbol{\beta}_{-p}^T\hat{\boldsymbol{\Sigma}}_{12} + \beta_p\hat{\Sigma}_{22} - 1\{p = i\} + \lambda\partial|\beta_p| = 0.$$

Thus the solution is the thresholding rule

$$\beta_p = \mathcal{T}\left(1\{p = i\} - \boldsymbol{\beta}_{-p}^T\hat{\boldsymbol{\Sigma}}_{12}, \lambda\right)/\hat{\Sigma}_{22}.$$

∎